\documentclass[reprint,footnoteinbib,prl,twocolumns]{revtex4-1}
\usepackage{amsfonts,amsmath,amssymb}
\usepackage[pdftex]{graphicx}
\usepackage{mathrsfs}
\usepackage{wasysym}
\usepackage{enumerate}
\usepackage{bm}

\usepackage[svgnames]{xcolor}
\usepackage{braket}
\usepackage[
    colorlinks=True,linkcolor=DarkRed,citecolor=ForestGreen,urlcolor=MediumBlue,
    pdfstartview=FitH,bookmarks=False,pdfpagemode=UseNone
]{hyperref}

\begin{document}

\title{Non-local nuclear spin quieting in quantum dot molecules: \\ Optically-induced extended two-electron spin coherence time}

\author{Colin M. Chow$^{1,2,*}$, Aaron M. Ross$^{1,*}$, Danny Kim$^3$, \\ Daniel Gammon$^3$, Allan S. Bracker$^3$, L. J. Sham$^4$, Duncan G. Steel$^{1,2,*}$ \\
\normalsize{$^1$H. M. Randall Laboratory of Physics, University of Michigan, Ann Arbor, Michigan} \\
\normalsize{$^2$Electrical Engineering and Computer Science, University of Michigan, Ann Arbor, Michigan} \\
\normalsize{$^3$Naval Research Laboratory, Washington D.C.} \\
\normalsize{$^4$Department of Physics, University of California San Diego, La Jolla, California} \\
\normalsize{$^*$These authors contributed equally to this work.}
}

\date{\today}

\begin{abstract}
We demonstrate the extension of coherence between all four two-electron spin ground states
of an InAs quantum dot molecule (QDM) via non-local suppression of nuclear spin fluctuations in both constituent quantum dots (QDs), while optically addressing only the upper QD transitions. Long coherence times are revealed through dark-state spectroscopy as resulting from nuclear spin locking mediated by the exchange interaction between the QDs. Lineshape analysis provides the first measurement of the quieting of the Overhauser field distribution correlating with reduced nuclear spin fluctuations.
\end{abstract}

\maketitle

The solid state community has made great strides in demonstrating basic quantum information operations using self-assembled quantum dot structures. Rapid, high-fidelity ground state initialization has been achieved in both electron\cite{Atature2006, Xu2007} and hole\cite{Brash2015} spin systems, ultrafast coherent spin rotations can be performed within picoseconds using detuned Raman pulses\cite{Press2008,Greilich2009,Kim2010}, spin states can be read out via absorption\cite{Kim2010} or fluorescence\cite{Press2008,Schaibley2013}, and spin-photon entanglement\cite{DeGreve2012,Gao2012,Schaibley2013} allows for the incorporation of these systems into larger quantum networks. Furthermore, the decoherence properties of both electron and hole spin qubits have been thoroughly investigated with data indicating that nuclear spin fluctuations via hyperfine coupling and residual charge fluctuations\cite{Kuhlmann2013,Stanley2014,Matthiesen2014,Urbaszek2013} are the primary sources of spin decoherence. Intense efforts have focused on how to protect the electron spin qubit from these noise sources, mainly resorting to dynamical decoupling using ultrafast pulses\cite{Press2010} and nuclear spin fluctuation quieting\cite{Xu2009,Sun2012}. In addition, recent experiments using optically-detected NMR revealed the advantages of strained InGaAs/GaAs QDs over strain-free GaAs/AlGaAs structures, predicting electron spin coherence times on the order of hundreds of microseconds\cite{Chekhovich2014}. 

The fabrication of high-quality vertically-stacked self-assembled quantum dot molecules (QDMs)\cite{Stinaff2006} has allowed the community to extend the studies discussed above to systems consisting of a few strongly-interacting electrons. Specifically, progress has been made in the coherent control of two electrons trapped in a QDM consisting of two quantum dots separated by a small tunneling barrier\cite{Doty2008,Robledo2008}. In such a two-electron system, a new qubit is formed from the singlet $S$ and the spin-projection zero triplet $T_0$ due to their relative insensitivity to charge and nuclear spin noise fluctuations. Ultrafast optical manipulation of the $S-T_0$ qubit has been demonstrated\cite{Kim2010a}, later followed by coherent population trapping (CPT) experiments, revealing coherence times of at least 200 ns\cite{Weiss2012}.

\begin{figure}
 \includegraphics[width=1.0\linewidth]{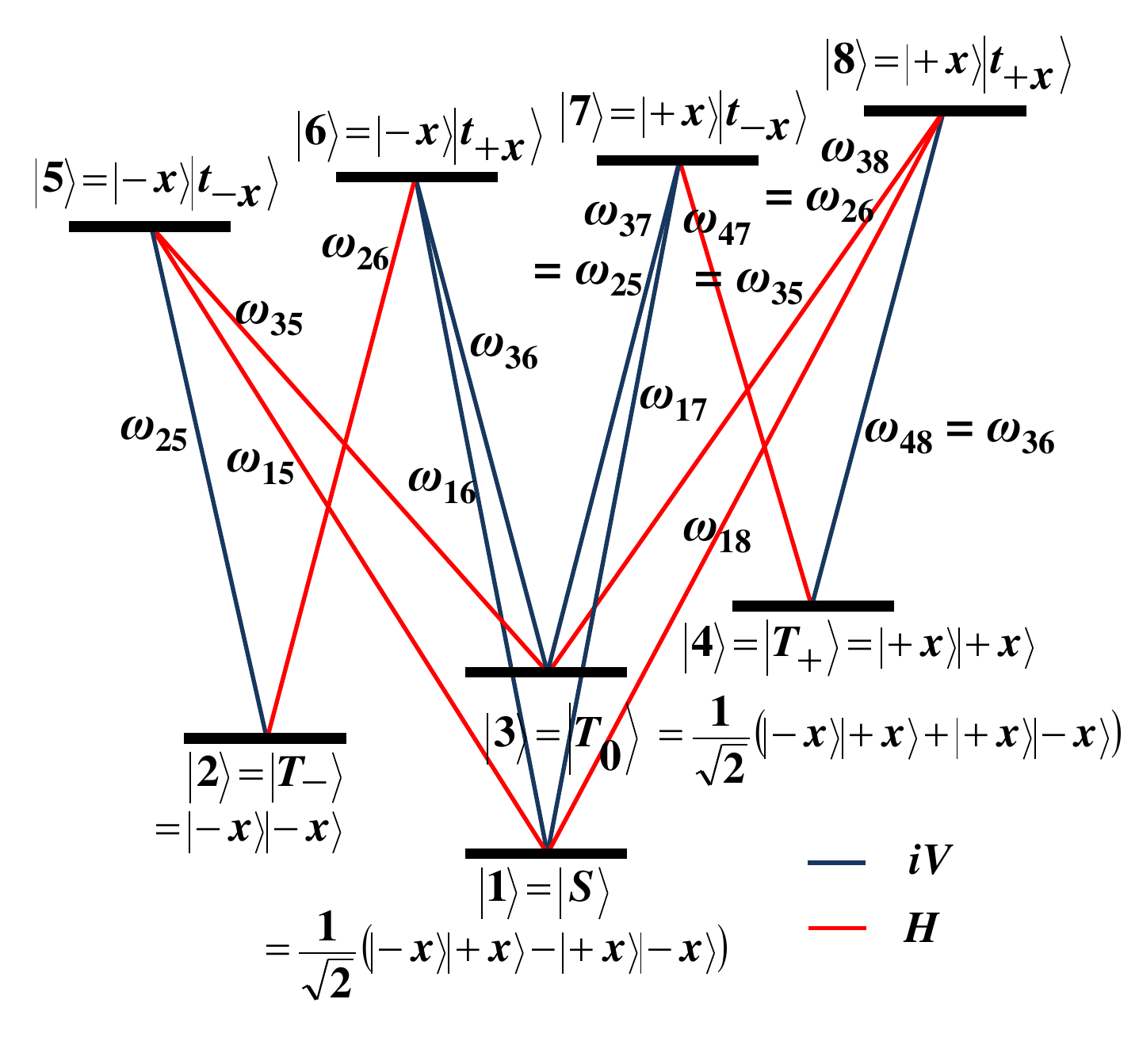}
  \caption{Energy level diagram and dipole-allowed optical transitions between singlet/triplet ground states and optically excited states in the presence of an in-plane (Voigt geometry) magnetic field. States consist of an electron in the lower QD, represented by $\ket{\pm x}$, and a trion in the upper QD, represented by $\ket{t_{\pm x}}$. Here, spin projections are shown along +x-direction and $+(-)$ denotes spin up (down). Blue (red) lines represent vertical (horizontal) polarization.}
 \label{fig:eight_level}
\end{figure}

\begin{figure*}
 \includegraphics[width=1.0\linewidth]{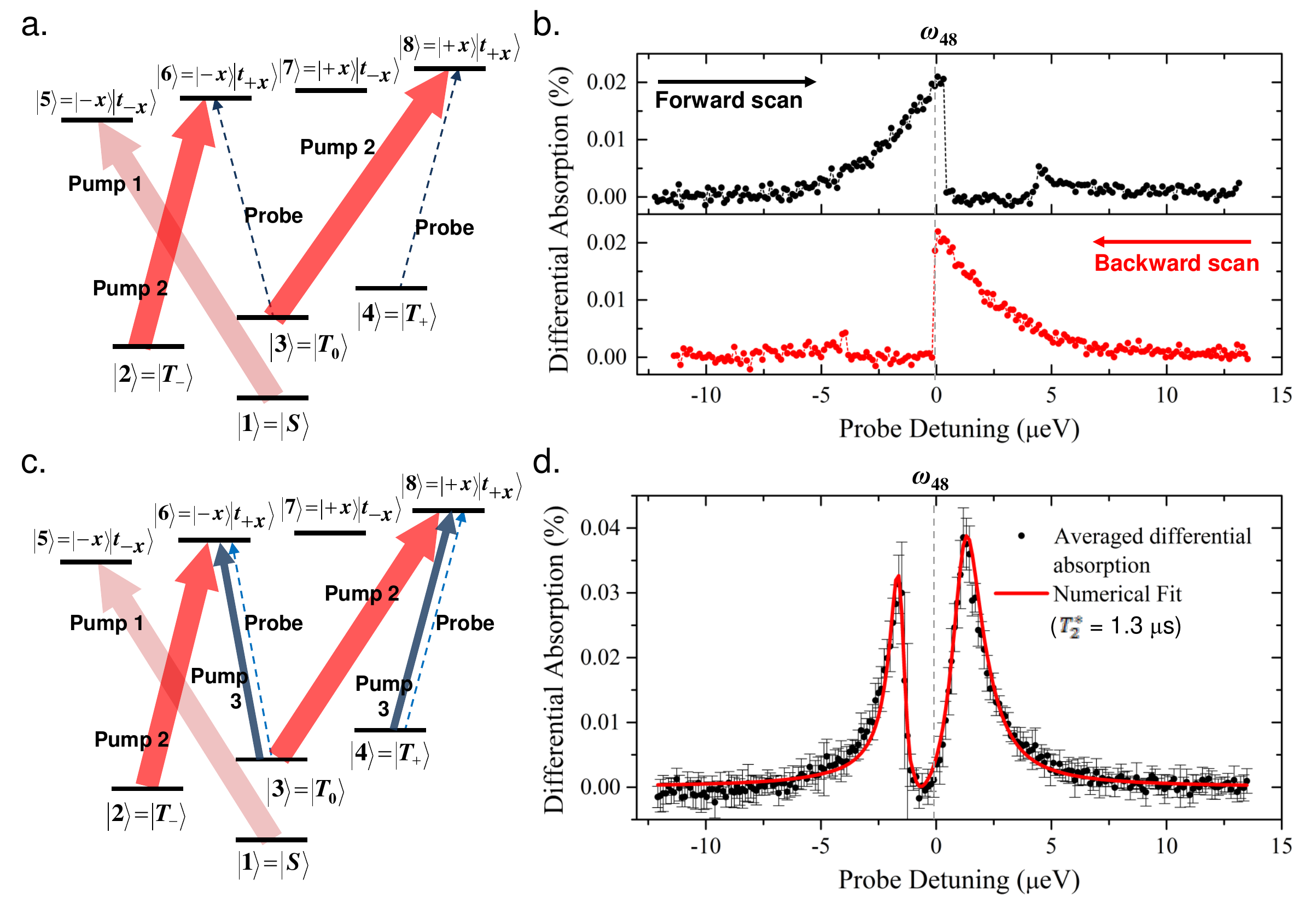}
  \caption{Suppression of DNSP by optical nuclear spin locking. a. Pump configuration for $T_+$ state preparation. b. Probe absorption spectra following the pumping scheme in a. In the upper panel, the vertically polarized probe is scanned in forward direction across $\omega_{48/36}$ transition. In the lower panel, the probe laser is scanned in backward direction. The spectra show hysteresis due to DNSP with the QD resonance seeming to move away from the approaching probe frequency. c. Pump 3 is added to the configuration shown in a to suppress the effect of DNSP. d. Probe absorption spectrum showing the recovery of dark-state profile. Solid circles in the plot represent averaged data points obtained from a series of 7 scans and the error bars show standard deviations. Red solid lines is the theoretical fit.}
 \label{fig:anti_dragging}
\end{figure*}

In this Letter, we demonstrate long coherence times for arbitrary superpositions of any of the four ground states ($S,T_0,T_{\pm}$) of the two-electron strongly-coupled QDM system, extending the QDM platform beyond a single $S-T_0$ system. These long coherence times are achieved via non-local suppression of nuclear spin fluctuations in both constituent quantum dots while optically addressing only one of the QDs of the molecule. Dark-state spectroscopy reveals that the nuclear spin fluctuation quieting is mediated by the strong exchange interaction between the two QDs. Analysis of the data provides the first measurement of the quieting of the Overhauser field distribution coinciding with the enhanced electron spin coherence time. We report a lower-bounded ground state coherence time of at least 1 microsecond.

In the set of experiments described below, we study an InAs/GaAs QDM consisting of two InAs quantum dots separated by a small tunneling barrier embedded within a Schottky diode \cite{Doty2008} (see Supplementary information for sample details). A focusing aspheric lens is mounted on the sample with its center aligned to a 6-$\mu$m-diameter aperture formed by the aluminum mask. A collimating aspheric lens is placed behind (on the substrate side) the sample. The assembly is inserted into a helium-flow cryostat operating at 6 K. All absorption spectra are measured using Stark shift modulation spectroscopy (see Supplementary Information). In all experiments, a magnetic field of 1.5 Tesla in Voigt geometry (perpendicular to the laser propagation) is applied. The g-factor for electron is found through magneto-absorption experiments to be 0.43 while for the heavy-hole, −0.084.

With an appropriate bias voltage, two electrons are confined in the QDM such that the wave function of each electron resides mostly in separate dots\cite{Doty2008}. The tunneling of each electron through the inter-dot barrier leads to the Heisenberg exchange interaction which forms the molecular states. Hence, the two-electron states consist of the spin-singlet, or $S$ state, and the triplet manifold, denoted by ${T_-, T_0, T_+}$, representing total spin down ($m_j$ = -1), zero ($m_j$ = 0) and up ($m_j$ = +1), respectively. With our choice of laser frequency, an electron-hole pair is created in the upper QD during optical excitation. The Coulomb interaction associated with this additional electron-hole pair shifts the relative electron energy levels between the QDs and prevents tunneling\cite{Doty2008}, thereby isolating the QDs. The optical excited states thus consist of a single electron in the lower QD and a trion, i.e., a heavy-hole and two spin-paired electrons, in the upper QD, with four possible spin configurations in which the spin state of the trion is given by its constituent heavy-hole.

The $S$ and $T_0$ states, with zero spin projection, form the decoherence-free subspace which is immune to fluctuations of the nuclear Overhauser field from the underlying lattice; these states therefore readily become the basis for a single qubit. Recent studies have demonstrated long spin coherence and ultrafast optical control of $S$-$T_0$ qubits in self-assembled InAs QDMs by exploiting the $\mathbf{\Lambda}$-system formed by dipole-allowed transitions between $S$-$T_0$ and a common excited state\cite{Kim2010a}. Ideally, in zero or an out-of-plane magnetic field (Faraday geometry), $T_-$ and $T_+$ states are decoupled from the $S$-$T_0$ subspace. Nonetheless, weak coupling may arise due to in-plane Overhauser field or heavy-hole-light-hole mixing\cite{Xu2009}, and could contribute to decoherence since $T_-$ and $T_+$ states are susceptible to fluctuations of Overhauser field. Alternatively, as shown in Fig. 1, an in-plane magnetic field (Voigt geometry) splits the triplets and allows $T_-$ and $T_+$ states to be coupled to the $S$-$T_0$ subspace via $\mathbf{\Lambda}$-systems. This offers the following advantages. First, ultrafast spin preparation of $T_-$ and $T_+$ states can be achieved with optical pumping, which establishes the first step towards utilizing QDMs for two-qubit operations. Second, as reported in the following, easy access to $T_-$ and $T_+$ states enables nuclear spin locking via hyperfine coupling between the electron and nuclear spins; \textit{this feature is not available to the $S$-$T_0$ qubit due to its intrinsic nuclear spin noise immunity. This results in the extension of two-qubit coherence by two orders of magnitude}.

\begin{figure*}
 \includegraphics[width=1.0\linewidth]{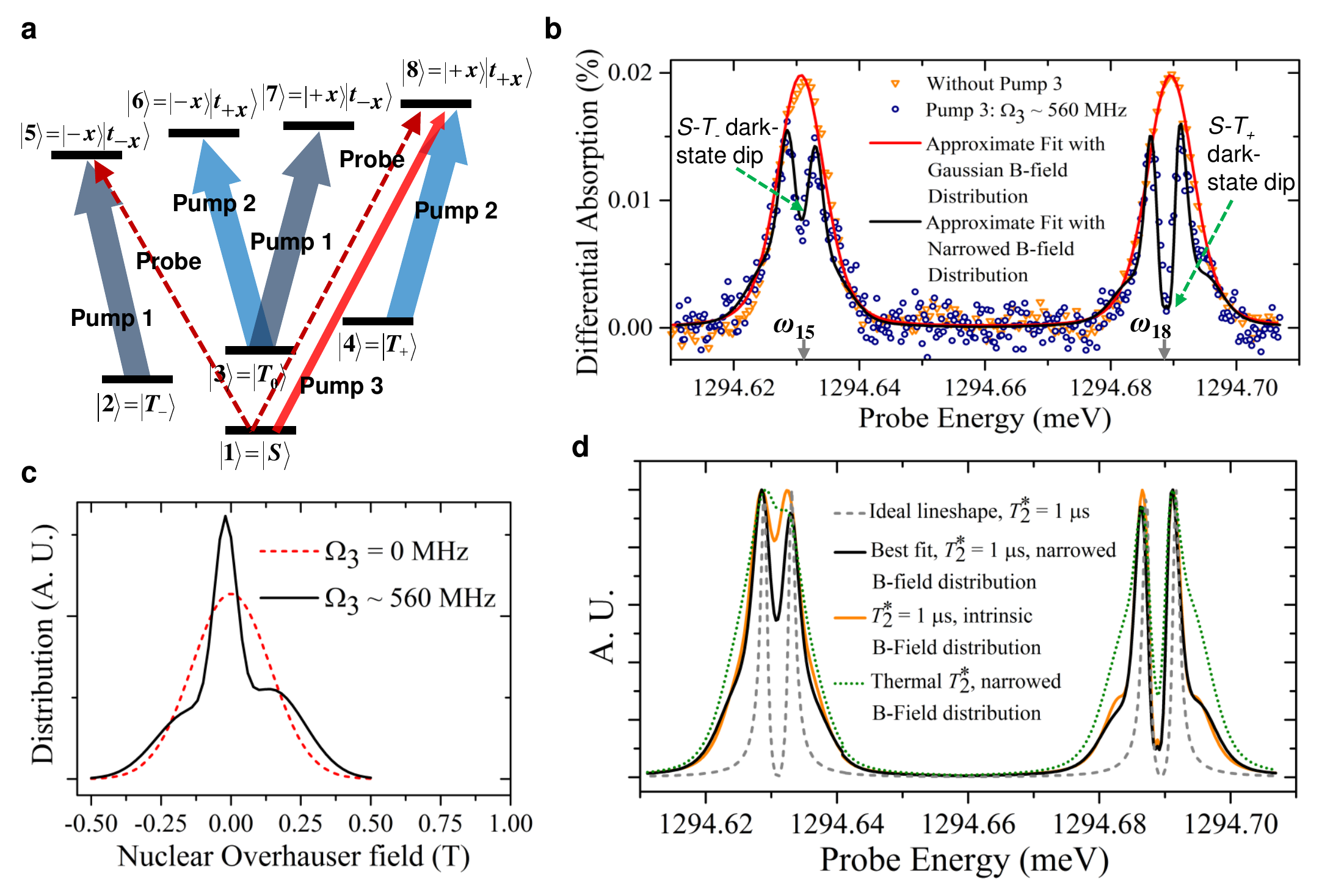}
  \caption{Singlet-triplet coherence and nuclear field distribution quieting. a, Pump configuration for preparation of a coherent $S$-$T_+$ superposition. b, Absorption spectra showing the emergence of dark-state dips from $S$-$T_-$ and $S$-$T_+$ coherence at transitions $\omega_{15}$ and $\omega_{18}$, respectively with the application of Pump 3. c, Nuclear field distributions used in the numerical model for fitting the spectra in b. The case without Pump 3 results in a broader Gaussian distribution (dotted curve) and the case with Pump 3 produces a narrower and more complex shape (solid curve). d, Comparison between spectra calculated from different combinations of decoherence times and Overhauser field distributions. }
 \label{fig:narrowing}
\end{figure*}

At 1.5 Tesla magnetic field in Voigt geometry, all Zeeman splittings (20 $\mu$eV) are larger than the optical transition linewidths (5 $\mu$eV) such that the incident laser (narrowband, $<$1 MHz) beams primarily drive transitions for which they are resonant. 

Of the twelve dipole-allowed transitions depicted in the eight-level system in Fig, 1, there are four pairs of energetically degenerate transitions, ($\omega_{25}, \omega_{37}$), ($\omega_{26}, \omega_{38}$), ($\omega_{35}, \omega_{47}$) and ($\omega_{36}, \omega_{48}$), arising from the fact that the electron g-factors of both QDs are nearly equal\cite{Greilich2013}. An out-of-equilibrium population of the $T_+$ state is achieved by the application of horizontally-polarized cw Pump 1 and 2, on resonance with transitions $\omega_{15}$ and $\omega_{26/38}$, respectively, as shown in Fig. 2a. By tuning the frequency of Pump 1 to $\omega_{15}$, where state 5 is not excited by another laser, coherent population trapping (CPT)\cite{Xu2008} is avoided and the system is prepared in $T_+$ state with near unity fidelity. This is confirmed by measuring the absorption of a vertically-polarized probe laser scanned across all four vertical transition frequencies, where a signal is observed only at $\omega_{48}$ and no absorption signal is measured at the other transitions (outside of displayed range) within the signal-to-noise ratio of the measurement. When the probe and Pump 2 form a two-photon resonance, i.e., when their frequency difference equals the Zeeman splitting of the triplets, CPT is expected, as demonstrated by a dark-state dip in the probe absorption spectrum. However, in both forward (increasing in frequency) and backward (decreasing in frequency) scans, the absorption spectra (Fig. 2b) show the more complex structures resembling distortion and hysteresis typical of dynamic nuclear spin polarization (DNSP). The spectra reveal the tuning of the optical resonance and, consequently, the two-photon resonance due to the shifting Overhauser field caused by the scanning of the probe. Here, the optical resonance moves away when the scanning probe frequency approaches, resulting in an abrupt reduction of the absorption signal. In the other case where the system is prepared in $T_-$ state, the resonance appears to follow the scanning of the probe, giving rise to an absorption profile with a round top, where resonant coupling is maintained over an extended range of about 10 $\mu$eV. (See Supplementary Information.)

To extract the decoherence time from the absorption profile, it is necessary to suppress the effect of DNSP due to the probe. Here, we apply a third pump laser, Pump 3, tuned to the two-photon resonance with Pump 2, as shown in Fig. 2c. A configuration for CPT is created in the $\bf{M}$-system as recently studied in atomic systems\cite{Gu2006} consisting of states 2, 6, 3, 8 and 4, such that a coherent state comprising all three triplet states is formed, with the probability amplitudes of individual states dictated by the relative intensities of the pumps and the relevant optical dipole moments. To the best of our knowledge, this is the first report of coherent control in a 5 level system in semiconductors and provides a novel platform for studies in electromagnetically-induced transparency. For the effective intensity ratio of Pump 2 to Pump 3 arbitrarily chosen to be 25 : 1, the system is coherently initialized to predominantly $T_+$ state. The Overhauser field is stabilized when the Rabi frequency of Pump 3 is adjusted to be sufficiently strong in order to overwhelm the dynamics otherwise induced by the scanning probe, thereby minimizing the perturbation on the nuclear spin caused by the probe. As revealed in Fig. 2d, a prominent dark state dip is now observed and the distortion in the lineshape is largely suppressed. This allows us to simulate the behavior of the system without considering the effects of DNSP by using the eight-level master equation (See Supplementary Information). The best-fit absorption lineshape is overlaid on the experimental result, in which the full depth of the dip indicates long ground state decoherence times, $T_2^*$, found to be at least 1.3 $\mu$s. This corresponds to an extension of $T_2^*$ by a factor of 500 compared to the estimated decoherence time due to thermally-distributed nuclear fluctuations\cite{Xu2009}, and therefore signifies a dramatic suppression of Overhauser field fluctuations in both QDs. \textit{The nuclear spin quieting is nonlocal due to the fact that optical excitations occur only in the upper QD.}

A unique feature of QDMs in contrast to single QDs is that the spin-zero singlet state enables the experimental study of the roles of $T_-$ and $T_+$ states in nuclear spin locking, as well as the direct observation of the associated quieting of Overhauser field distribution. In the pump configuration shown in Fig. 3a, the system is prepared in a coherent superposition of the singlet and $T_+$ state when Pump 2 and Pump 3 are in two-photon resonance. In the measured probe absorption spectrum (hollow circles in Fig. 3b), a dark-state dip is seen in each of the singlet transitions, as expected from the CPT. Nonetheless, the lineshapes deviate from the ideal dark-state profile, while the depths of the dark-state dips in both transitions increase as the intensity of Pump 3 is raised. Without Pump 3, however, the system is prepared in the singlet state and the dark-state dips vanish (triangles in Fig. 3b) in the now broadened lineshapes, contrary to what is expected from CPT in a $\Lambda$-system, assuming an extended coherence time between the ground states. This can be explained by considering a stochastic effective magnetic environment due to fluctuations in the nuclear spins of the underlying lattice. Although the singlet state is unaffected by the Overhauser field, the fluctuating Overhauser field affects both the spin-polarized upper-QD trion, and the $T_-$ and $T_+$ states, via the Zeeman shift, thus making the two-photon resonance condition unstable and obscuring any dark state dip. To construct a theoretical fit, we assume a spectral diffusion model where the Overhauser field is assumed to be slowly varying compared to optical processes\cite{Wang1991}. The calculated absorption spectra corresponding to different individual Overhauser fields are then averaged according to the best-fitting Overhauser field distribution. Here the intrinsic Overhauser field (the case without Pump 3) follows a Gaussian distribution with an extracted standard deviation of 0.15 Tesla (dashed line in Fig. 3c), in agreement with the theoretical order-of-magnitude estimate of 0.11 Tesla (See Supplementary Information). The resulting lineshape, shown as the red solid line in Fig. 3b, suggests that the averaging of different spectra is sufficient to obscure the dark-state dip as observed in both simulation and experiment, without invoking enhanced nuclear spin fluctuations.

Remarkably, when Pump 3 is applied, the same fitting procedure produces a narrowed distribution of the Overhauser field, as shown by the solid line in Fig. 3c, and qualitatively reproduces the observed dark-state profiles of the two singlet transitions simultaneously. The appearance of both dark-state dips signifies a long spin decoherence time between the optical ground states, here estimated to be about 1 $\mu$s, consistent with that reported earlier. This interpretation is corroborated by our simulations with different scenarios as described in Fig. 3d. In particular, for two limiting cases where in the first, an intrinsic Overhauser field distribution together with a long $T_2^*$ of $\mu$s is assumed, and in the second, a narrowed Overhauser field distribution with a thermal spin decoherence time of 2.5 ns\cite{Xu2009}, neither of the resulting lineshapes fits the data. Our model also accounts for the difference in the depths of the dips, where the dark-state dip at $\omega_{15}$ is shallower due to the finite width of the Overhauser field, while at $\omega_{18}$, the dip is enhanced by Pump 3 which saturates the optical transition. When the Overhauser field distribution follows a Dirac delta function, both transitions show a full dark-state dip. Hence, the data presented in Fig. 3b along with line shape analysis provides an experimental means to determine the Overhauser field quieting following optically induced nuclear spin polarization via a nonzero $T_+$ population.

In conclusion, we have demonstrated new degrees of freedom for coherently manipulating the electronic spin states and the underlying nuclear spin ensemble in a coupled-QD system. In particular, it is possible to stabilize the Overhauser field through the application of appropriate optical fields. The important new physics includes the observation that the nuclear spin quieting leads to ground state decoherence times in excess of one microsecond and extends over both QDs, even though the driving of the hyperfine coupling via the optical excitation is localized to the upper QD. Since nuclear spin diffusion across the tunnel barrier between QDs is unlikely due to strain and energy mismatch in Zeeman shift\cite{Maletinsky2009,Chekhovich2012,Chekhovich2014,Urbaszek2013}, this non-local nature likely stems from electron-mediated nuclear spin flip-flop\cite{Urbaszek2013,Deng2005,Huang2010,Latta2011} between two spatially separated QDs. The underlying mechanisms of DNSP in QDMs are yet to be fully elucidated. Nonetheless, our observation of long spin coherence involving all four two-electron spin eigenstates suggests that InAs QDMs can potentially be used to implement multiple two-qubit gates within its $T_2^*$ time. Based on recent observation that nuclear spin polarization in single QDs persists beyond one second, and in special cases out to 30 hours\cite{Latta2011}, it is reasonable to expect similar result from QDMs. If this is the case, the hyperfine interaction between confined electrons and nuclei in QDMs may be exploited for potential applications such as quantum memory\cite{Taylor2003}, or as a platform for the fundamental study of mesoscopic entanglement between two nuclear ensembles\cite{Schuetz2014}. Our work sets the stage for developing a fully scalable architecture for quantum information processing using optically accessible “hard-wired” QD complexes, now on par with recent accomplishments in the trapped ion and gate-defined coupled quantum dot communities.

  \bibliography{nuclear}
\end{document}